\documentclass[twocolumn, secnumarabic, amssymb, nobibnotes, aps, prl,showpacs,superscriptaddress]{revtex4-1}
\usepackage{graphicx}
\usepackage{float}
\usepackage{bm}

\usepackage{bbm}
\newcommand{\equref}[1]{Eq.~(\ref{#1})}

\newcommand{\figref}[1]{Fig.~\ref{#1}}
\newcommand{\refcite}[1]{Ref.~\onlinecite{#1}}
\newcommand{\refscite}[1]{Refs.~\onlinecite{#1}}

\newcommand{\tableref}[1]{Table~\ref{#1}}

\renewcommand{\vec}[1]{\boldsymbol{#1}}
\usepackage{color}

\usepackage[colorlinks=true]{hyperref}
\hypersetup{
    bookmarks=true,         % show bookmarks bar?
    unicode=false,          % non-Latin characters
    pdftoolbar=true,        % show Acrobat
    pdfmenubar=true,        % show Acrobat
    pdffitwindow=false,     % window fit to page when opened
    pdfstartview={FitH},    % fits the width of the page to the window
    pdftitle={Full gap superconductivity from London penetration depth measurements in type-II Dirac semimetal PdTe2},    % title
    pdfauthor={Teknowijoyo et al.},     % author
    pdfsubject={},   % subject of the document
    pdfcreator={},   % creator of the document
    pdfproducer={}, % producer of the document
    pdfkeywords={} {} {}, % list of keywords
    pdfnewwindow=true,      % links in new window
    colorlinks=true,       % false: boxed links; true: colored links
    linkcolor=blue, %red,          % color of internal links (change box color with linkbordercolor)
    citecolor=blue,        % color of links to bibliography
    filecolor=magenta,      % color of file links
    urlcolor=blue           % color of external links
}

\begin{document}

\title{Nodeless superconductivity in type-II Dirac semimetal PdTe$_2$:  low-temperature London penetration depth and symmetry analysis}

\author{Serafim Teknowijoyo}
%\email{seraphtw@iastate.edu}
\affiliation{Ames Laboratory, Ames, Iowa 50011, USA}
\affiliation{Department of Physics and Astronomy, Iowa State University, Ames, Iowa 50011, USA }	

\author{Na Hyun Jo}
%\email{njo@iastate.edu}
\affiliation{Ames Laboratory, Ames, Iowa 50011, USA}
\affiliation{Department of Physics and Astronomy, Iowa State University, Ames, Iowa 50011, USA }

\author{Mathias S. Scheurer}
%\email{mscheurer@g.harvard.edu}
\affiliation{Department of Physics, Harvard University, Cambridge, Massachusetts 02138, USA}

\author{M.~A.~Tanatar}
%\email{tanatar@ameslab.gov}
\affiliation{Ames Laboratory, Ames, Iowa 50011, USA}
\affiliation{Department of Physics and Astronomy, Iowa State University, Ames, Iowa 50011, USA}

\author{Kyuil Cho}
\affiliation{Ames Laboratory, Ames, Iowa 50011, USA}

\author{S. L. Bud'ko}
%\email{budko@ameslab.gov}
\affiliation{Ames Laboratory, Ames, Iowa 50011, USA}
\affiliation{Department of Physics and Astronomy, Iowa State University, Ames, Iowa 50011,
USA }

\author{Peter P.~Orth}
%\email{porth@iastate.edu}
\affiliation{Department of Physics and Astronomy, Iowa State University, Ames, Iowa 50011,
USA }

\author{P.~C.~Canfield}
%\email{canfield@ameslab.gov}
\affiliation{Ames Laboratory, Ames, Iowa 50011, USA}
\affiliation{Department of Physics and Astronomy, Iowa State University, Ames, Iowa 50011,
USA }

\author{R.~Prozorov}
\email{Corresponding author: prozorov@ameslab.gov}
\affiliation{Ames Laboratory, Ames, Iowa 50011, USA}
\affiliation{Department of Physics and Astronomy, Iowa State University, Ames, Iowa 50011,
USA }

\date{Submitted: 1 April 2018}

\begin{abstract}
Superconducting gap structure was probed in type-II Dirac semimetal PdTe$_2$ by measuring the London penetration depth using tunnel diode resonator technique. At low temperatures, the data for two samples are well described by weak coupling exponential fit yielding $\lambda(T=0)=230$~nm as the only fit parameter at a fixed $\Delta(0)/T_c\approx 1.76$, and the calculated superfluid density is consistent with a fully gapped superconducting state characterized by a single gap scale. Electrical resistivity measurements for in-plane and inter-plane current directions find very low and nearly temperature-independent normal- state anisotropy. The temperature dependence of resistivity is typical for conventional phonon scattering in metals. We compare these experimental results with expectations from a detailed theoretical symmetry analysis and reduce the number of possible superconducting pairing states in PdTe$_2$ to only three nodeless candidates: a regular, topologically trivial, $s$-wave pairing, and two distinct odd-parity triplet states that both can be topologically non-trivial depending on the microscopic interactions driving the superconducting instability.

\end{abstract}

%74.70.Xa Pnictides and chalcogenides
%74.25.Dw Superconductivity phase diagrams
%72.15.-v Electroniup-trianglesc conduction in metals and alloys

\maketitle

\section{Introduction}
Finding materials that exhibit topological superconductivity is one of the primary goals of current research efforts in condensed matter physics, mainly motivated by their unique Majorana surface state properties~\cite{QiZhang, HasanKane, Ando}. These protected non-Abelian surface modes~\cite{ReadGreen, Kitaev-1} can, for example, be exploited in quantum computing schemes~\cite{Nayak}. The search for topological superconductors (TSCs) has recently been boosted by the discovery of various material classes that feature topological band structures already in the normal state. There are multiple scenarios in which superconducting pairing among such states, characterized by a non-zero topological invariant (\emph{e.g.} a Chern number), result in the emergence of topological superconductivity~\cite{QiZhang, HasanKane, Ando}.

Examples are topological insulators, which feature non-degenerate two-dimensional (2D) Dirac surface cones. Superconductivity arises either from doping such as in Cu$_x$Bi$_2$Se$_3$~\cite{FuBerg, Hor-2010, Maeno} and Sb$_2$Te$_3$~\cite{Zhao-Sb2Te3}, or from proximity-coupling of the 2D Dirac surface state to a regular $s$-wave SC~\cite{FuKane-2008, HasanKane}. Other examples are semiconductor heterostructures and quantum wires with strong spin-orbit coupling and (proximity-induced) superconductivity~\cite{Beenaker-Review}. Three-dimensional examples are magnetic, \emph{i.e.}, inversion-symmetric, Weyl semimetals (SM) that favor odd-parity (often topological) pairing over ordinary even-parity pairing~\cite{Ando, Cho-2012, Burkov-2015}. This is a result of the unique spin texture on the Fermi surfaces surrounding the Weyl points~\cite{Ando, Armitage-Review-2018}. In addition, time-reversal invariant Weyl SM were shown to host topological superconductivity for suitable electronic interactions~\cite{Hosur}. Three-dimensional Dirac SM such as Cd$_2$As$_3$~\cite{He-Cd2As3}, Na$_3$Bi~\cite{Liu-Na3Bi} (type-I) and PdTe$_2$~\cite{Noh-2017} (type-II) are proper starting points to realize Weyl SMs by either breaking inversion or time-reversal symmetry, \emph{e.g.}, via magnetic order or external fields. Moreover, Dirac SMs have been predicted to be a rich platform for topological (crystalline) SC themselves, at least for the $C_4$ symmetric systems Cd$_2$As$_3$ and Au$_2$Pb~\cite{Sato-PRL-2015, Ando}.

Here, we investigate superconductivity in single-crystals of the transition metal dichalcogenide PdTe$_2$ (space group P$\bar{3}m1$), which is a type-II Dirac SM~\cite{Noh-2017, dHvA,PdTe-ARPES, manipulation}. As shown by ARPES and band structure calculations~\cite{PdTe-ARPES, dHvA}, the Dirac band crossing occurs about 0.6~eV below the Fermi energy and is protected by $C_3$ rotation symmetry. In addition, quantum oscillation measurements of the de Haas-van Alphen effect~\cite{dHvA} revealed a non-zero Berry phase in one of the (hole) Fermi surface pockets, confirming the topological nature of the band crossing. Notably, the superconducting state that we study emerges below 1.7~K~\cite{Bell,Canadian}.

We report experimental results of the London penetration depth using a tunnel diode resonator (TDR) technique~\cite{vandegrift}. Our findings clearly indicate a fully gapped superconducting state, in agreement with previous thermodynamic~\cite{type1SC}, scanning tunnel microscopy (STM)~\cite{STM}, and heat capacity~\cite{HC} measurements. Combining these experimental insights with a detailed theoretical symmetry analysis, we are able to reduce the possible superconducting pairing states in PdTe$_2$ to only three candidates, two of which can be topologically non-trivial.

The remaining candidate SC pairing states are the standard $s$-wave BCS state, which is topologically trivial, and two time-reversal symmetric odd-parity triplet states transforming under the representations $A_{1u}$ and $E_g$ of the point group $D_{3d}$ of the normal state. Whether these odd-parity states are topologically trivial or non-trivial, depends on the relative sign of the superconducting order parameter on the two Fermi surfaces enclosing the $\Gamma$ point in the Brillouin zone and, hence, is determined by whether the electron-electron interactions between the different pockets is repulsive or attractive.
We suggest that experiments which are able to systematically tune the impurity scattering rate, for example \emph{via} electron irradiation, could be used to further distinguish between the three remaining SC states.

\section{Experimental details}

\begin{figure}
	\includegraphics[scale=1]{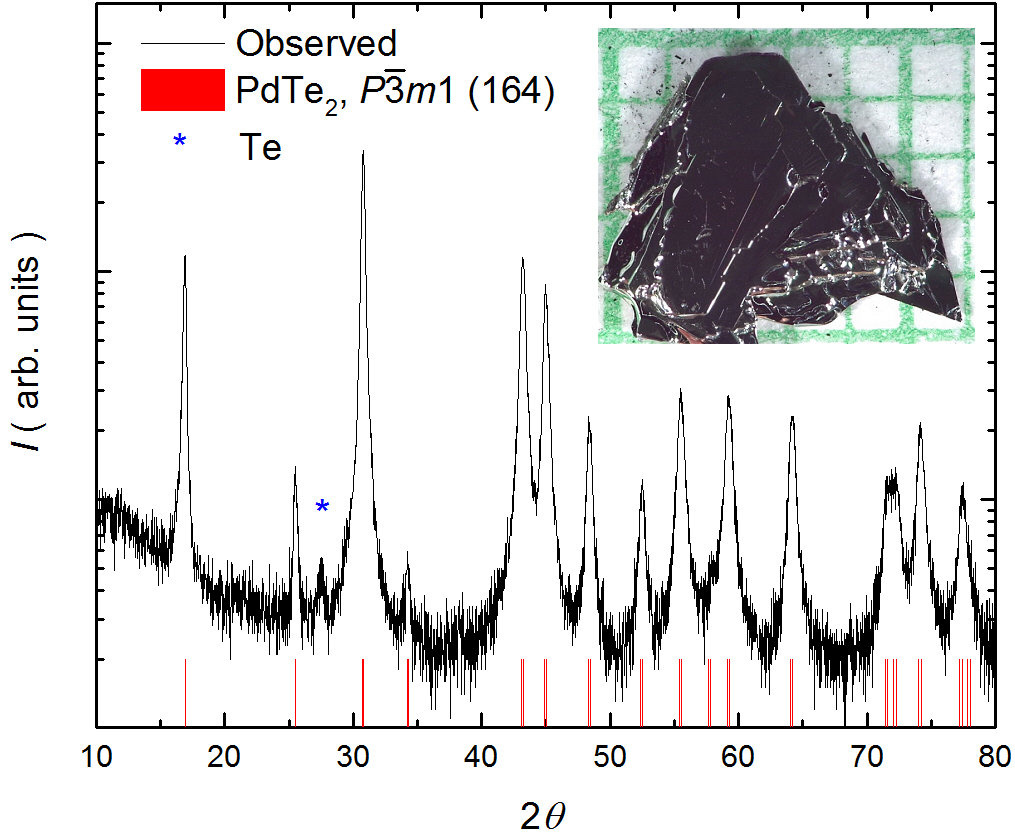}%
	\caption{(Color online) Powder x-ray diffraction (XRD) pattern of crashed single crystal of PdTe$_{2}$ (black line). The red lines are calculated XRD peaks for PdTe$_{2}$ with hexagonal structure[$P\bar{3}$m1, 164]. Blue stars mark peaks of solidified Te flux.
		\label{xrd}}
\end{figure}

\begin{figure}[tb]
\begin{center}
\includegraphics[width=0.90\linewidth]{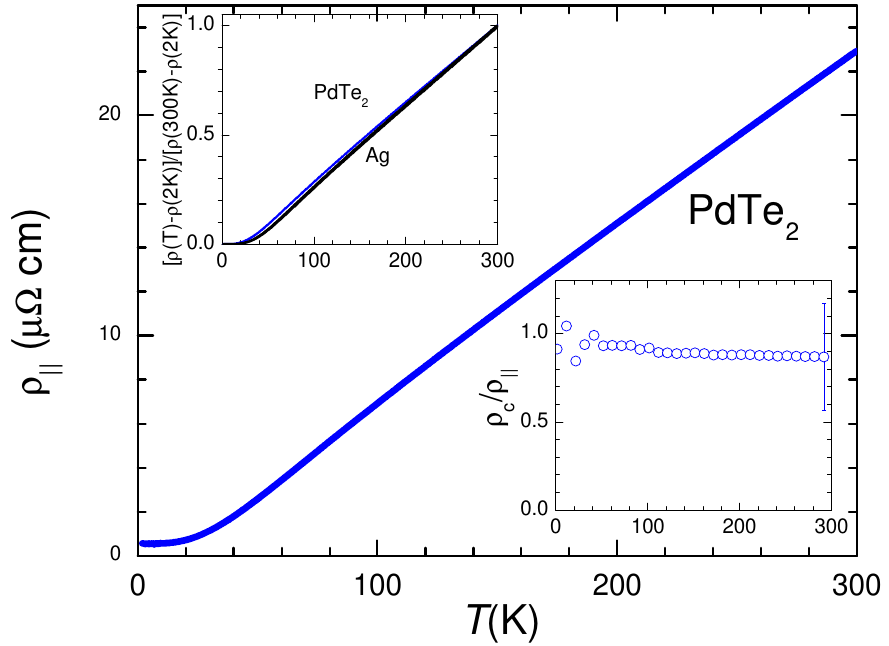}
\end{center}
\caption{Temperature-dependent in-plane electrical resistivity of PdTe$_2$. Note a range below approximately 10~K where resistivity becomes temperature-independent and linear increase above 40~K. Left inset compares normalized temperature-dependent part of resistivity in PdTe$_2$ and Ag wire \cite{MTC}.  Right inset shows temperature-dependent resistivity anisotropy ratio, $\rho_c/\rho_{\parallel}$, with error bars determined by uncertainty of experimental geometry.
}%
\label{resistivity}
\end{figure}

Single crystals of PdTe$_2$ were grown out of Te rich binary melts. Elemental Pd(99.9+$\%$) and Te(Alfa Aesar, 99.999+$\%$) were put into a Canfield Crucible Set (CCS)~\cite{Canfield2016} with initial stoichiometry, Pd$_{0.10}$Te$_{0.90}$, and sealed in an amorphous silica tube. The ampules were heated up to 900$^\circ$~C, within 10 hours, held for 5 hours, cooled to 500$^\circ$~C, over 120 hours, and finally decanted using a centrifuge \cite{Canfield1992}. The obtained single crystals of PdTe$_{2}$ were hexagonal plate in morphology as shown in Fig.\,\ref{xrd} inset.

 Rigaku MiniFlex II diffractometer (Cu $K_{\alpha}$ radiation with monochromator) was used for acquiring a powder x-ray diffraction (XRD) pattern at room temperature. The acquired patterns are well matched with calculated peaks for hexagonal structure of PdTe$_{2}$ with $P\bar{3}$m1 (164) as shown in Fig.\,\ref{xrd}. Small intensity extra peak marked with blue star is associated with residual Te solvent left on the crystals. However, the relative intensity of the peaks is different from the calculated powder pattern, presumably because of the preferential orientation of the ground powder due to the layered structure.

Samples used for four-probe in-plane  electrical resistivity, $\rho_{\parallel}$, measurements were cleaved from inner parts of large single crystals and had dimensions of typically (2-3)$\times$0.5$\times$0.1 mm$^3$ with longer side along an arbitrary direction in hexagonal plane. Silver wires were soldered using In to the fresh-cleaved surface of the samples \cite{FeSedetwinning} to make electrical contacts with sub-m$\Omega$ resistance.  Sample resistivity at room temperature, $\rho(300K)$, was determined as $\rho(300K)=$24 $\pm$5 $\mu \Omega$cm, as determined on array of 7 samples.  This is consistent with early report \cite{Canadian} but is notably lower than 70 $\mu \Omega$cm reported recently \cite{HC}. Montgomery technique \cite{Montgomery,Montgomery2} measurements were performed on a sample with 1 mm by 0.5 mm cross-section area in the $ac$ plane of the crystal. Contacts were soldered on sample corners covering the whole length of the sample in the third dimension. Large uncertainty of geometric factor in the crystal due to non-negligible contact size (typically 0.1 mm) compared to the sample size make these measurements semi-quantitative. The anisotropy value $\rho_c/\rho_{\parallel}=$0.9$\pm$0.3 was found to be temperature independent, see Fig.~\ref{resistivity}. Temperature dependent electrical resistivity measurements in four-probe and Montgomery configurations were performed down to 1.8~K in {\it Quantum Design} PPMS.

Precision in-plane London penetration depth $\Delta\lambda(T)$ measurements using TDR technique \cite{vandegrift} were performed in a high stability  $^3$He-cryostat with the base temperature of $\sim$0.4 K. Two samples $\#$A and $\#$B were measured. The samples were placed with their $c$-axis parallel to an excitation field, $H_{ac} \sim 20$ mOe, much smaller than $H_{c1}$~\cite{type1SC}. The shift of the resonant frequency, $\Delta f(T)=-G4\pi\chi(T)$, is proportional to the differential magnetic susceptibility $\chi(T)$. The constant $G=f_0V_s/2V_c(1-N)$ depends on the demagnetization factor $N$, sample volume $V_s$ and coil volume $V_c$. $G$ was determined from the full frequency change by physically pulling the sample out of the coil. With the characteristic sample size, $R$, $4\pi\chi=(\lambda/R)\tanh (R/\lambda)-1$, from which $\Delta \lambda$ can be obtained \cite{Prozorov2000,Prozorov2006}.

\section{Results}

The main panel of Fig.~\ref{resistivity} shows temperature dependent in-plane resistivity of PdTe$_2$. Despite relatively high resistivity value $\rho(300K)=24$~$\mu \Omega$cm, the dependence is very typical of a good metal: it is $T$-linear for $T\gtrsim $~K, and flattens below approximately 10~K in the residual resistivity range before the superconducting transition at $T_c\sim 1.7$~K (not shown). Direct comparison of the temperature-dependent part of resistivity, $[\rho(T)-\rho(0)]/[\rho(300K)-\rho(0)]$,  with that of Ag wire \cite{MTC} is made in the left top inset in Fig.~\ref{resistivity} and finds a nearly perfect match. Slightly lower end of $T$-linear range in PdTe$_2$ is caused by slightly lower Debye temperature, $\Theta_D\sim 207$~K \cite{Debay} as compared with 225 K in Ag. This observation clearly identifies phonon scattering as the main scattering mechanism. Nearly isotropic resistivity without noticeable temperature dependence (right bottom inset in Fig.~\ref{resistivity}) identifies the material as being three-dimensional, in agreement with band structure calculations~\cite{FS}.

\begin{figure}[tb]
\begin{center}
\includegraphics[width=0.90\linewidth]{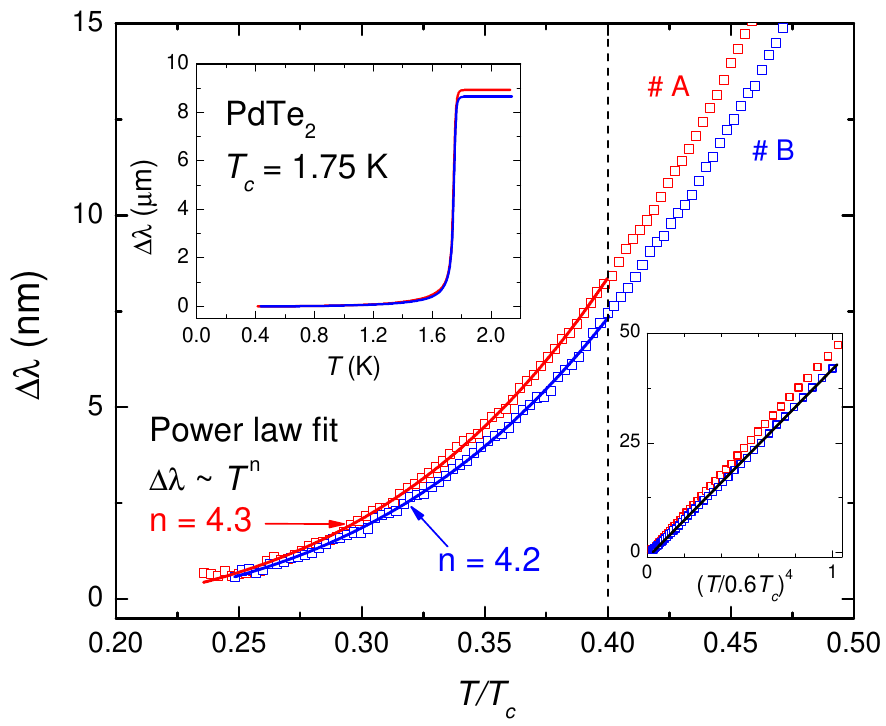}
\end{center}
\caption{(Color online) Temperature variation of London penetration depth $\Delta \lambda (T)$ measured in He$^3$ TDR setup for samples $\#$A (red) and $\#$B (blue). Main panel shows data with best fit using power-law function $\Delta \lambda (T)=A+BT^n$, with $n=$4.3 ($\#$A) and $n=$4.2 ($\#$B).Right bottom panel shows same data plotted as a function of $T^4$ to verify quality of the fit. Top inset shows data over the whole temperature range up to $T_c\sim$1.8~K.
}%
\label{DeltaLambda}
\end{figure}

\begin{figure}[tb]
\begin{center}
\includegraphics[width=0.90\linewidth]{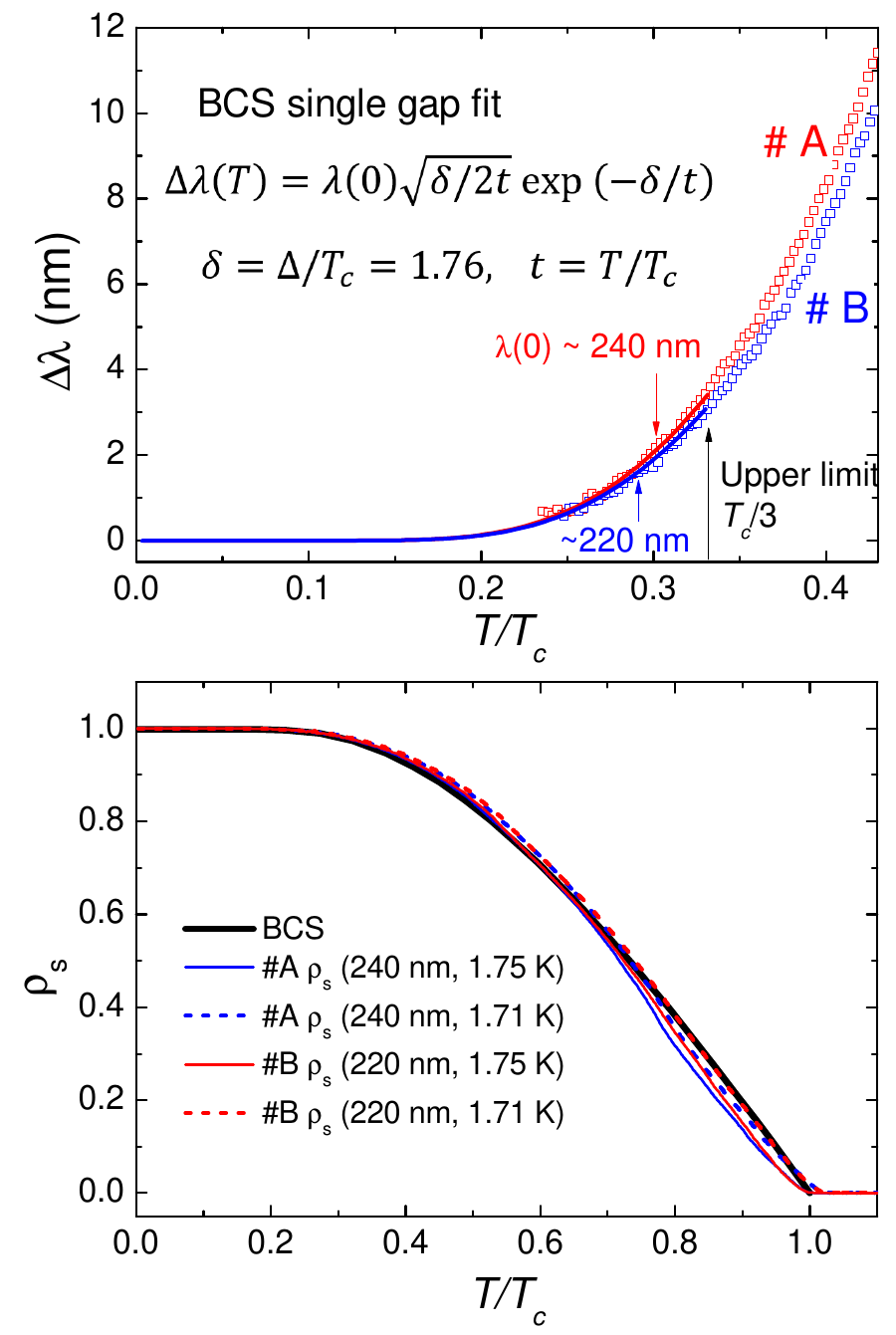}
\end{center}
\caption{(Color online) Top panel. Fit of the temperature variation of London penetration depth with exponential function, enabling determination of $\lambda (0)$=240 nm (sample $\#$A, red) and $\lambda (0)$=220 nm (sample $\#$B, blue). Bottom panel shows calculated superfluid density $\rho_s(T)\equiv (\lambda(0)/\lambda(T))^2$ assuming various values of $T_c$.
}%
\label{rhos}
\end{figure}

Left top inset in Fig.~\ref{DeltaLambda} shows temperature dependent penetration depth in PdTe$_2$, measured over the whole range of superconductivity existence. The superconducting transition with $T_c=1.75$~K is very sharp, as expected in stoichiometric materials. The main panel of Figure \ref{DeltaLambda} shows low-temperature part of temperature variation of $\Delta\lambda(T)$ in two single crystalline samples ($\#$A, red, and $\#$B, blue) of PdTe$_2$.  The data are shown on a normalized temperature scale $T/T_c$ in a temperature range below $0.5T_c$. In the clean limit, the temperature-dependent London penetration depth is expected to be exponential in full gap superconductors and is expected to be close to $T$-linear is superconductors with nodes in the gap. Addition of  sufficiently strong disorder pushes the dependence to $T^2$ for both cases \cite{ProzorovKogan2011}. We use a power-law function $\Delta \lambda (T)=A+BT^n$ to quantify the experimental data for the intermediate cases, when the amount of disorder is not known. Note that the gap magnitude can vary either on the same Fermi surface sheet (gap anisotropy) or between different sheets of the Fermi surface (multi-band superconductivity).

It is empirically accepted that a variation described by the  power-law function with $n>$3 corresponds to the case of a full gap, and $n<$ 2 corresponds to a nodal case. This fit is made in a characteristic range below $0.3T_c$, in which the temperature dependence of the superconducting gap magnitude is negligible in single gap superconductors, and the dependence is determined by thermal excitation of quasi-particles across the superconducting gap. The red (blue) line in the main panel of Fig.~\ref{DeltaLambda} shows our power-law fit of the data for sample $\#$A ($\#$B) of PdTe$_2$ over the range up to $0.4T_c$, which yields the exponent $n=$4.3 ($n=$4.2). To check the quality of the fit in the bottom right inset in Fig.~\ref{DeltaLambda} we plot the penetration depth data as a function of $(T/T_c)^4$, finding a close to linear dependence for both samples.

A power-law function with such a large value of the exponent ($n \approx 4$) is indistinguishable from an exponential function (over the range of temperatures observed), which is the expected behavior for penetration depth in a fully gapped BCS superconductors \cite{BCS}. We therefore also fit our data using an exponential temperature dependence of $\Delta \lambda$. In the top panel of Fig.~\ref{rhos} we show the resulting fit of the London penetration depth data using the regular BCS expression. We obtain a good fit of the penetration depth using the zero temperature value
\begin{equation}
\lambda(0)=240~\text{nm (sample $\#$A)}
\end{equation}
and
\begin{equation}
\lambda(0)=220~\text{nm (sample $\#$B).}
\end{equation}
The determination of $\lambda(0)$ is very important, since the tunnel diode resonator technique does not enable to measure $\lambda(0)$ directly. The values we determined are consistent between samples giving average $\lambda(0)= 203$nm and are notably different from the estimate $\lambda(0)= 39$~nm based on Hall effect carrier density \cite{type1SC}. The origin of the discrepancy may potentially lie in the compensated character of Hall transport in PdTe$_2$ \cite{dHvA}, leading to an overestimate of the carrier density. Our experimentally determined value of $\lambda(0)$  may suggest notably higher value of Ginzburg-Landau parameter $\kappa$ than suggested in Ref.~\onlinecite{type1SC} and type-II superconductivity. Additional measurements are clearly needed to clarify this important question. Thermal conductivity may be a good candidate, since it allows to distinguish between bulk and surface superconducting states and the normal state via the Wiedemann-Franz law~\cite{WF}. It can also be used to distinguish between first and second order phase transitions~\cite{thermalcond1stvs2nd}.

Experimentally determined $\lambda(0)$ allows us to construct the temperature-dependent normalized superfluid density as $\rho_s=(\lambda(0)/\lambda(T))^2$, with $\lambda(T)=\lambda(0)+\Delta \lambda (T)$. In the bottom panel of Fig.~\ref{rhos} we show the resulting superfluid density $\rho_s$, calculated using our experimental data and the values $\lambda(0)$ as determined from the exponential fit. The data are plotted versus temperature $T/T_c$ (normalized to $T_c$), and compared with BCS expectations for a single fully gapped superconductor (dashes). There is some uncertainty in this plot, since the exact value of $T_c$ depends on the criterion used for its determination (onset versus maximum derivative in the top left panel of Fig.~\ref{DeltaLambda}). Both values give curves that lie close to expectations for BCS full-gap superconductors. This clearly shows that superconductivity in PdTe$_2$ is characterized by a single and full superconducting gap.

\begin{table*}[t]
\begin{center}
\caption{Possible pairing states in PdTe$_2$ as constrained by the point group $D_{3d}$. We use $X$, $Y$, and $Z$ to represent real-valued continuous functions on the Brillouin zone with the same transformation properties under $D_{3d}$ as $k_x$, $k_y$, and $k_z$. Here $a$, $b$, and $c$ are real coefficients that are not fixed by symmetry and follow from microscopic details of the system. The column TRS indicates whether time-reversal symmetry is preserved (y) or broken (n). The last three columns show the form of the order parameter using the pseudospin basis (see main text), the minimal number of nodes on a Fermi surface enclosing the $\Gamma$ point, and, for the fully gapped states, whether the phase is necessarily topologically trivial or can be topological depending on microscopic details.}
\label{PossiblePairingStates}
 \begin{tabular} {ccccccc} \hline \hline
    group th. & pairing  & $d_n$  & TRS & Order parameter $\Delta i\sigma_y$ &Minimal $\#$ nodes per FS & \hspace{1em}Topology \hspace{1em}\\ \hline
 $A_{1g}$ & $s$-wave  &  $1$ & y & $a+b (X^2+Y^2)+cZ^2$ & $0$ & \hspace{1em} trivial   \\
$A_{2g}$ & $g$-wave   &  $1$ & y & $XZ(X^2-3Y^2)$  & $4$ nodal lines &\hspace{1em} --- \\ \hline
$E_g$ & $e_{g(1,0)}$-wave   &  $2$ & y & $a (X^2-Y^2) + b YZ$ & $2$ nodal lines &\hspace{1em}--- \\
$E_g$ & $e_{g(0,1)}$-wave   &  $2$ & y & $a XY + b XZ$ & $2$ nodal lines &\hspace{1em}---  \\
$E_g$ & $e_{g(1,i)}$-wave   &  $2$ & n & $a (X+iY)^2 + b Z(Y+iX)$ & $2$ nodal points & \hspace{1em} ---  \\ \hline
$A_{1u}$ & $p$-wave   & $1$ & y & $a (X\sigma_x+Y\sigma_y)+b Z \sigma_z$ & $0$& \hspace{1em} trivial/top.   \\
$A_{2u}$  & $p$-wave   & $1$ & y & $a(Y\sigma_x - X\sigma_y)+bX(X^2-3Y^2)\sigma_z$ & $2$ nodal points & \hspace{1em} --- \\ \hline
$E_u$  & $e_{u(1,0)}$-wave   & $2$ & y & $a X(X^2-3Y^2) \sigma_x + b Z \sigma_y + c Y \sigma_z$ & $0$ & \hspace{1em}trivial/top. \\
$E_u$  & $e_{u(0,1)}$-wave   & $2$ & y & $a Z \sigma_x + b X(X^2-3Y^2) \sigma_y + c  X\sigma_z$ & $2$ nodal points & \hspace{1em}---  \\
$E_u$  & $e_{u(1,i)}$-wave   & $2$ & n & \hspace{1em} $[a Z + i b X(X^2-3Y^2)] (\sigma_x+i\sigma_y) + c (X+iY) \sigma_z $ & $2$ nodal points \cite{Comment} &\hspace{1em}---  \\ \hline \hline
 \end{tabular}
\end{center}
\end{table*}

\section{Discussion}
In the following, we will discuss the implications of our experimental findings for the possible superconducting order parameters.
Focusing on superconducting phases that do not break lattice translation symmetry, we can classify different pairing states according to the irreducible representations (IRs) of the point group $D_{3d} = \bar{3}\frac{2}{m}$ of the normal state of PdTe$_2$.
The resulting $10$ possible pairing states are summarized in \tableref{PossiblePairingStates}; four states arise from the four one-dimensional (1D) IRs ($d_n=1$) and three from each of the two 2D IRs ($d_n=2$). Here we choose the coordinate system such that $k_z$ refers to the $c$ direction, while $k_x$ and $k_y$ are momenta in the $ab$-plane with $k_x$ pointing along one of the two-fold rotation axes of $D_{3d}$ perpendicular to the $c$ direction.

%In general, the superconducting order parameter is a matrix $\Delta_{\alpha\beta}(\vec{k})$ with $\alpha$ and $\beta$ referring not only to spin, but also to several atomic orbitals .
To give explicit expressions for the microscopic form of the different order parameters in \tableref{PossiblePairingStates}, we use the pseudospin basis: Although spin is not a good quantum number in the presence of spin-orbit coupling (and several relevant orbitals), we can still define a ($\vec{k}$-space local) pseudospin basis with the same transformation properties as spin if the system has time-reversal and inversion symmetry. As long as different bands do not come close to each other, we can focus on a single band for a given $\vec{k}$-point and, hence, restrict the superconducting order parameter $\Delta(\vec{k})$ to be a $2\times 2$ matrix in pseudospin space. As usual, we expand this matrix in (pseudospin) singlet, $\psi$, and (pseudospin) triplet, with triplet vector $\vec{d}$, \emph{i.e.},
\begin{equation}
\Delta(\vec{k}) = \left( \sigma_0 \psi(\vec{k}) + \vec{d}(\vec{k}) \cdot \vec{\sigma} \right) i\sigma_y\,,
\end{equation}
where $\sigma_j$, $j=x,y,z$, denote Pauli matrices and $\sigma_0$ the identity matrix in pseudospin space. Due to the presence of inversion symmetry, all pairing channels in \tableref{PossiblePairingStates} are either pure singlet (gerade IRs) or triplet (ungerade IRs).

From \refscite{PdTe-ARPES, dHvA}, we know that there are two Fermi surfaces enclosing the $\Gamma$ point.
For this reason, we have analyzed the minimal number of nodal points or lines the different pairing states have on a Fermi surface that encloses the $\Gamma$ point.
From the result summarized in \tableref{PossiblePairingStates}, we can see that $7$ out of the $10$ pairing states will necessarily give rise to nodal lines or points and are, hence, inconsistent with our penetration depth measurements that clearly indicate a fully established gap on all Fermi surfaces. Consequently, only three pairing states remain possible -- the $s$-wave singlet state, the $p$-wave order parameter transforming under $A_{1u}$, and the $e_{u(1,0)}$ state.

Due to the preserved time-reversal symmetry, all of the remaining candidate pairing states belong to symmetry class DIII which is characterized by a $\mathbbm{Z}$ topological invariant $\nu$ in three spatial dimensions \cite{QiZhang}. To analyze $\nu$, let us first focus on one of the bands enclosing the $\Gamma$ point. In the case of the $s$-wave singlet state, we just have the standard BCS $s$-wave superconductor that is known to be topologically trivial. In the limit where the separation between the different bands at the Fermi level is larger than the superconducting order parameter, the invariant $\nu$ of the full system is given by the sum of the invariants $\nu_n$ of the different Fermi surfaces $n$, \emph{i.e.}, $\nu = \sum_n \nu_n$ \cite{Invariants}. For the $s$-wave singlet state, we just have $\nu_n =0$ on all Fermi surfaces $n$ and, hence, a trivial state $\nu=0$, irrespective of the relative phases of the order parameter on the different bands.

This is different for the $A_{1u}$ state: Focusing for the moment on the leading terms of the basis functions in the vicinity of the $\Gamma$ point, $X \sim k_x$, $Y\sim k_y$, and $Z\sim k_z$, the corresponding triplet vector reads $\vec{d}(\vec{k}) \sim (a k_x,a k_y, bk_z)$.
For just a single Fermi surface enclosing the $\Gamma$ point, we thus have an anisotropic form of the Balian-Werthamer state of the B phase of superfluid $^3$He. This state is known to be topologically non-trivial with $|\nu| = 1$~\cite{QiZhang,Volovik}.
Taking into account higher order terms in $X,Y,Z$, the invariant of the single Fermi surface can be different but must always be odd and, hence, nontrivial. This follows from the general result of \refcite{FuBerg} stating that the parity of the invariant $\nu$ of a superconducting order parameter that is odd under inversion is given by the parity of the number $N$ of the time-reversal invariant momenta ($\vec{k}=-\vec{k}$) enclosed by the Fermi surfaces of the system (which is one in the present case with one Fermi surface around the $\Gamma$ point), $\nu \, \text{mod}\, 2 = N \, \text{mod}\, 2$. %Note this even holds when different bands overlap and the pseudospin description is not valid any more.

\begin{figure}[tb]
\begin{center}
\includegraphics[width=\linewidth]{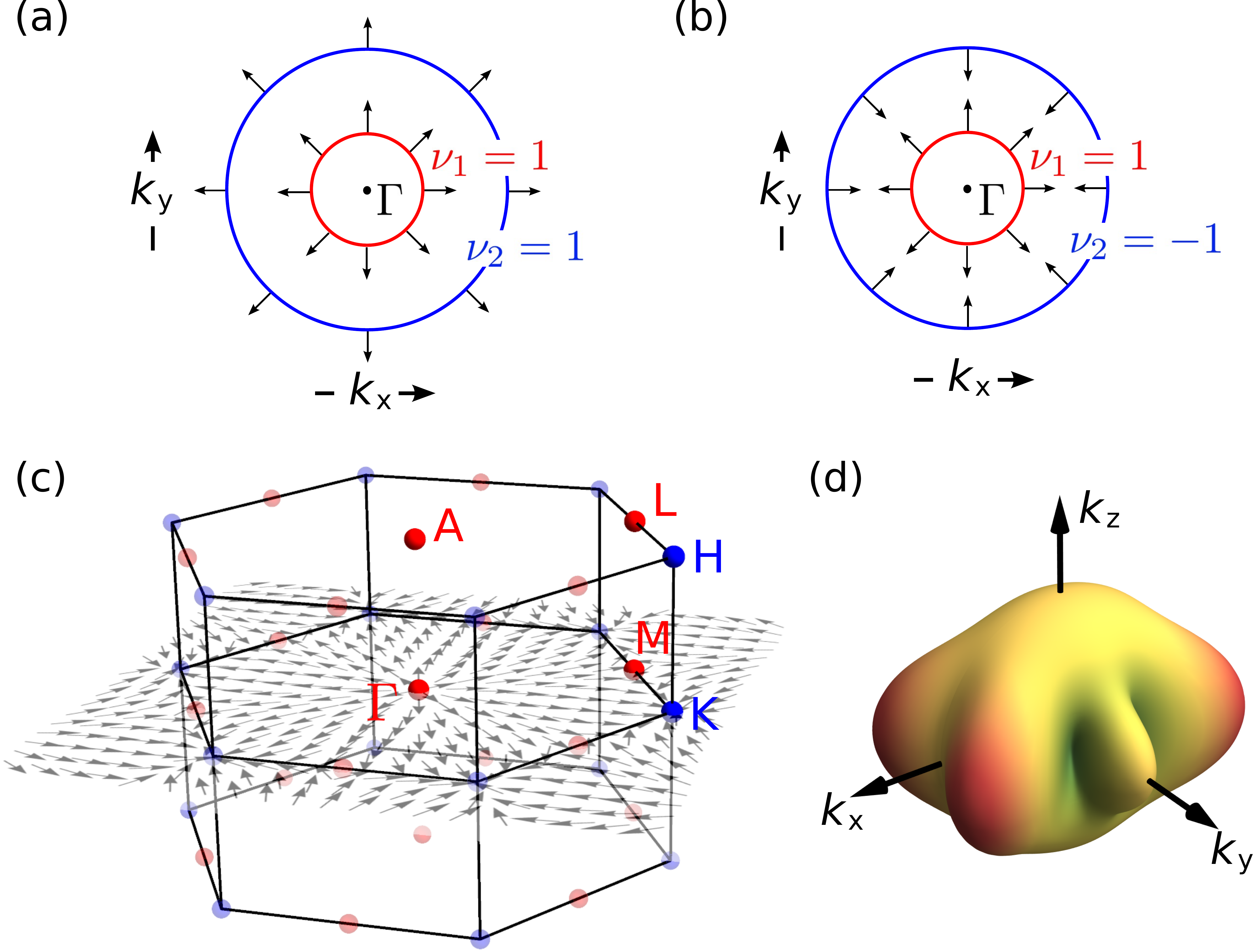}
\end{center}
\caption{(Color online) The triplet vector (black arrows) on the two Fermi surfaces (red and blue solid lines) enclosing the $\Gamma$ point is shown in (a), for the same sign of the order parameter on the two Fermi surfaces, and (b), for opposite signs, yielding a topologically non-trivial ($\nu_1+\nu_2\neq 0$) and trival state ($\nu_1+\nu_2= 0$), respectively. In (c), we indicate the high-symmetry points, where the triplet vectors of the two candidate states, $A_{1u}$ and $e_{u(1,0)}$ in \tableref{PossiblePairingStates}, have to vanish as a consequence of inversion symmetry (red dots) and rotation symmetries (blue dots). The gray arrows illustrate the (simplest) texture of the triplet vector of the $A_{1u}$ state in the $k_z=0$ plane (with minimal number of defects). Part (d) shows the directional dependence (\emph{i.e.}, anisotropy) of the gap of the $e_{u(1,0)}$ state on a Fermi surface enclosing the $\Gamma$ point.  The distance of the surface to the origin is proportional to the magnitude of the gap. The state breaks the three-fold rotation symmetry and its gap is, thus, generically anisotropic.
}%
\label{TheoryPlots}
\end{figure}

Unfortunately, we cannot apply the criterion for topological superconductivity of \refcite{FuBerg} to the invariant $\nu$ of the full system as the total number of enclosed time-reversal momenta is even \cite{PdTe-ARPES, dHvA}. In other words, the interplay between different bands that are topological individually determines whether $\nu=0$ or $\nu\neq0$. E.g, if there is no additional sign change of the triplet order parameter between the two Fermi surfaces enclosing the $\Gamma$ point, the invariants of the two bands add to the non-trivial value $\nu = 2$, see \figref{TheoryPlots}(a). On the other hand, in the presence of an additional sign change, the invariants cancel, resulting in a trivial state, $\nu=0$ as illustrated in \figref{TheoryPlots}(b). In general, the situation is more complicated due to the presence of additional pockets \cite{PdTe-ARPES, dHvA} away from the $\Gamma$ point. \emph{E.g.}, the predicted pockets around the K and K' points are generically expected to yield a non-trivial contribution to $\nu$ although K, K' are not time-reversal invariant. The reason is that the K and K' points are high-symmetry points where the triplet vector of the $A_{1u}$ state is forced to vanish due to rotational symmetry [see \figref{TheoryPlots}(c)]. For the $A_{1u}$ state, symmetry also enforces that the contributions $\nu_{\text{K}}$ and $\nu_{\text{K'}}$ of the Fermi surfaces enclosing the K and K' point to the invariant are equal, \emph{i.e.}, $\nu_{\text{K}} = \nu_{\text{K'}}$. %In contrast, they may contribute differently in case of the $E_u$ state.

While both pairing states discussed so far have a gap that is invariant under all symmetry operations of the normal state, the third candidate, the $e_{u(1,0)}$ state, transforms as $k_x$ under $D_{3d}$ and, hence, has a gap that breaks the three-fold rotation symmetry along the $c$-axis, see \figref{TheoryPlots}(d).
As can be easily seen by adiabatic deformation (see appendix), choosing the lowest order basis functions, $X \sim k_x$, $Y\sim k_y$, and $Z\sim k_z$, again gives $|\nu|=1$ for a single Fermi surface enclosing the $\Gamma$ point.
In fact, the previous discussion of the topological invariant of the $A_{1u}$ state based on \refcite{FuBerg} equally well apply to the $e_{u(1,0)}$-wave order parameter. The only difference is that there is in general no relation for the $e_{u(1,0)}$ state between the topological invariants $\nu_{\text{K}}$ and $\nu_{\text{K}'}$ of the Fermi surfaces enclosing the K and K' points as the three-fold rotation symmetry along $k_z$ is broken.

As indicated in the last column of \tableref{PossiblePairingStates}, this shows that, while the $s$-wave phase is a topologically trivial state, both of the odd parity candidate phases can be either topologically trivial or nontrivial depending on microscopic details.

\section{Conclusions and Outlook}
\label{sec:conclusions}
We have presented measurements of the London penetration depth using TDR technique and of the resistivity in single-crystals of type-II Dirac semimetal PdTe$_2$. Our results reveal that the SC state is fully gapped and characterized by a single gap energy scale. This is in agreement with previous STM, magnetization and AC susceptibility results. We determine a zero temperature London penetration depth of $\lambda(0) \approx 230$~nm from a fit of our measurements of $\Delta \lambda(T)$. Combining this with the previously measured value of $\xi = 114$~nm~\cite{type1SC}, one finds $\kappa = \lambda/\xi \approx 2.0$. This is slightly larger than $1/\sqrt{2} = 0.7$, corresponding to type-II superconductivity. However, in view of rather convincing thermodynamic evidence for type-I~ superconductivity \cite{type1SC}, this question deserves further investigation, \emph{e.g.}, using thermal conductivity measurements. We also report a temperature-dependence of the resisitivity and its anisotropy that do not reveal any anomalous features and instead closely following expectations for an isotropic metal with dominant phonon scattering.

%Despite observation of Dirac fermions in PdTe$_2$, the temperature-dependent resistivity of the material does not reveal any anomalous features and closely follows expectations for isotropic metal with dominant phonon scattering. The superconducting state emerging in this metal has full single gap, consistent with STM measurements. Our measurements do not allow us to confirm or disprove type-I superconducting state in the material.
%Questions: Do we need the last sentence? Is this generally impossible with TDR measurements or is there a suble reason for it here?

We have performed a systematic theoretical analysis of all possible SC pairing states that can be reached by a single continuous phase transition from the normal state.
Using as input our results of a full superconducting gap together with the known form of the Fermi surfaces~\cite{PdTe-ARPES, dHvA}, we are able to narrow down the possible SC pairing states to only three candidates: An $s$-wave superconductor transforming trivially under all symmetries of the lattice, a $p$-wave phase transforming under $A_{1u}$, and a triplet order parameter ($e_{u(1,0)}$) transforming as $k_x$ under $D_{3d}$.

While the first state is always topologically trivial, the latter two triplet phases can be topologically non-trivial, depending on the relative sign of the SC order parameter on different Fermi surfaces. The crucial difference between the triplet states is that the gap of the $A_{1u}$ order parameter is invariant under all lattice symmetries, whereas the gap of the  $e_{u(1,0)}$ state breaks the three-fold rotation symmetry along the $c$ axis of the normal state.

While our transport measurements indicate the relevance of phonons for momentum relaxation, it is not clear whether phonons also provide the paring glue.
This is important as electron-phonon coupling alone is expected to yield a topologically trivial state, even in the (time-reversal symmetric) Weyl SM state that can be reached by adding an inversion-symmetry-breaking perturbation~\cite{Brydon,Scheurer-2016}. The situation is different for magnetic Weyl SMs, which preserve inversion symmetry. Here, the singlet $s$-wave pairing state is not allowed due to the spin structure around the Weyl points and the pairing state necessarily has odd-parity~\cite{Ando}. Alternatively, adding magnetic impurities may, in principle, also result in topological superconductivity~\cite{Scheurer-2016}. Further microscopic calculations are necessary to understand the connection between the interplay of different electron-electron interaction channels and the resulting superconducting order parameter.

Finally, to experimentally distinguish between the remaining candidate states, we suggest to investigate the different behavior of the SC transition temperature $T_c$ when tuning the impurity scattering rate, \emph{e.g.} via electron irradiation.

\section{Acknowledgements}
The authors would like to thank Morgan Masters, Joshua Slagle and Victor Barrena Escolar for support during the crystal growth. The experimental work was supported by the U.S. Department of Energy (DOE), Office of Basic Energy Sciences, Division of Materials Sciences and Engineering. The experimental research was performed at Ames Laboratory,  which is operated for the U.S. DOE by Iowa State University under Contract No.~DE-AC02-07CH11358.
N.H.J. is supported by the Gordon and Betty Moore Foundation's EPiQS Initiative (Grant No. GBMF4411).
M.S.S. acknowledges support from the German National Academy of Sciences Leopoldina through grant LPDS 2016-12. P.P.O. acknowledges support from Iowa State University Startup Funds.

%%%%%%%%%%%%%%%%%%%%%%%%%%%% BIBLIOGRAPHY

%\vspace{2em}

%\newpage

%==================================================================================================================================
\appendix
\section*{Adiabatic deformation of the ${e_{u(1,0)}}$ state}
For completeness, we here present a simple argument showing that the two candidate triplet states, $A_{1u}$ and $e_{u(1,0)}$ in \tableref{PossiblePairingStates} with leading order basis functions around the $\Gamma$ point ($X\sim k_x$, $Y\sim k_y$, $Z\sim k_z$), are topologically equivalent, \emph{i.e.}, have the same topological invariant $\nu$. To this end, let us define the set of triplet vectors
\begin{equation}
\vec{d}_{\eta}(\vec{k}) = \left(a (1-\eta)k_x(k_x^2-3k_y^2) + a\eta k_x,b k_z,ck_y\right),
\end{equation}
which can be used to interpolate between the $e_{u(1,0)}$ state, at $\eta=0$, and
\begin{equation}
\vec{d}_{\eta=1}(\vec{k}) = \left(a k_x, b k_z, c k_y\right). \label{DeformedState}
\end{equation}
It is easily seen that $|\vec{d}_{\eta}(\vec{k})|\neq 0$ for $0 \leq \eta \leq 1$, $\vec{k}\neq 0$. Consequently, the gap does not close which guarantees that $\nu$ does not change during the deformation. Performing a rotation in spin space, which again keeps the gap intact and does not affect $\nu$, the triplet vector in \equref{DeformedState} can be deformed continuously into $-\left(ak_x,ck_y,bk_z\right)$. This is the form of the triplet vector of the $A_{1u}$ order parameter, which proves the topological equivalence of the two states.


\begin{thebibliography}{44}%

\bibitem{QiZhang} X.-L.~Qi and S.-C.~Zhang,
Rev. Mod. Phys. {\bf 83}, 1057 (2011).
%Topological insulators and superconductors

\bibitem{HasanKane} M. Z.~Hasan and C. L.~Kane, Rev. Mod. Phys. \textbf{82}, 3045 (2010).

\bibitem{Ando} M.~Sato and Y.~Ando, Rep. Prog. Phys. {\bf 80}, 076501 (2017).
% Review on topological superconductors

\bibitem{ReadGreen} N.~Read, D.~Green, Phys. Rev. B. \textbf{61}, 10267 (2000).
%p-wave SC with non-Abelian anyons

\bibitem{Kitaev-1} A.~Kitaev, Phys. Usp. \textbf{44}, 131 (2001).
%Unpaired Majoranas in quantum wires

\bibitem{Nayak} C.~Nayak, S. H.~Simon, A.~Stern, M.~Freedman, and S.~Das~Sarma, Rev. Mod. Phys. \textbf{80}, 1083 (2008).
% Review on topological QC

\bibitem{FuBerg} L. Fu and E. Berg, Phys. Rev. Lett. {\bf 105}, 097001 (2010).
% Proposal for odd-parity pairing in CuxBi2Se3

\bibitem{Hor-2010} Y.S. Hor, A. J. Williams, J. G. Checkelsky, P. Roushan, J. Seo,
Q. Xu, H. W. Zandbergen, A. Yazdani, N. P. Ong and R. J. Cava, Phys. Rev.
Lett. \textbf{104}, 057001 (2010).
% Experimental realization of SC in CuxBi2Se3

\bibitem{Maeno} S. Yonezawa, K. Tajiri, S. Nakata, Y. Nagai, Z. Wang, K. Segawa, Y. Ando, and Y. Maeno, Nat. Phys. {\bf 13}, 123 (2017).
%Thermodynamic evidence for nematic superconductivity in CuxBi2Se3

\bibitem{Zhao-Sb2Te3} L. Zhao, H. Deng, I. Korzhovska, M. Begliarbekov, Z. Chen, E. Andrade,
E. Rosenthal, A. Pasupathy, V. Oganesyan, and L. Krusin-Elbaum, Nat. Comm. \textbf{6}, 8279 (2015).


\bibitem{FuKane-2008} L.~Fu and C. L.~Kane, Phys. Rev. Lett. \textbf{100} 096407 (2008).
% s-wave proximity induced topological SC on surface of TI

\bibitem{Beenaker-Review} C. W. J.~Beenaker, Annu. Rev. Condens. Matter Phys. \textbf{4} 113 (2013).
%Beenaker Review on Majorana fermions in SOC heterostructures


\bibitem{Cho-2012} G. Y. Cho, J. H. Bardarson, Y.-M. Lu, and J. E. Moore, Phys. Rev. B \textbf{86}, 214514 (2012).
% Topo SC in doped inversion symmetric Weyls

\bibitem{Burkov-2015} G. Bednik, A. A. Zyuzin, and A. A. Burkov, Phys. Rev. B \textbf{92}, 035153 (2015).
% Topo SC in inversion-symmetric Weyl SM

\bibitem{Armitage-Review-2018} N. P. Armitage, E. J. Mele, and A. Vishwanath
Rev. Mod. Phys. \textbf{90}, 015001 (2018).


\bibitem{Hosur} P.~Hosur, X.~Dai, Z.~Fang, and X.~-L.~Qi, Phys. Rev. B \textbf{90}, 045130 (2014).
% Topo SC in doped time-reversal invariant Weyl SM

\bibitem{He-Cd2As3} L. P. He, Y. T. Jia, S. J. Zhang, X. C. Hong, C. Q. Jin and S. Y. Li, NPJ Quantum Mater. \textbf{1} 16014 (2016).


\bibitem{Liu-Na3Bi} Z. K. Liu, B. Zhou, Y. Zhang, Z. J. Wang, H. M. Weng, D. Prabhakaran, S.-K. Mo, Z. X. Shen, Z. Fang, X. Dai, Z. Hussain, Y. L. Chen, Science  \textbf{343}, 864 (2014).


\bibitem{Noh-2017}
H.-J. Noh, J. Jeong, E.-J. Cho, K. Kim, B. I. Min, and B.-G. Park, Phys. Rev. Lett. {\bf 119}, 016401 (2017).
%Experimental Realization of Type-II Dirac Fermions in a PdTe2 Superconductor

\bibitem{Sato-PRL-2015} S. Kobayashi and M. Sato, Phys. Rev. Lett. \textbf{115} 187001 (2015).

\bibitem{dHvA}
F. Fei, X. Bo, R. Wang, B. Wu, J. Jiang, D. Fu, M. Gao, H. Zheng, Y. Chen,
X. Wang, H. Bu, F. Song, X. Wan, B. Wang, and G. Wang, Phys. Rev. B {\bf 96}, 041201(R) (2017).
%Nontrivial Berry phase and type-II Dirac transport in the layered material PdTe2

\bibitem{PdTe-ARPES} Y. Liu , J. Z. Zhao, L. Yu, C.-T. Lin, A.-J. Liang, C. Hu, Y. Ding, Y. Xu, S.-L. He, L. Zhao, G.-D. Liu, X.-L. Dong, J. Zhang, C.-T. Chen, Z.-Y. Xu, H.-M. Weng, X. Dai, Z. Fang, X.-J. Zhou, Chin. Phys. Lett. \textbf{32}, 067303 (2015).
% ARPES on PdTe2

\bibitem{manipulation} R. C. Xiao, P. L. Gong, Q. S. Wu, W. J. Lu, M. J. Wei, J. Y. Li, H. Y. Lv, X. Luo, P. Tong, X. B. Zhu, and Y. P. Sun, Phys. Rev. B {\bf 96}, 075101 (2017).
%Manipulation of type-I and type-II Dirac points in PdTe2 superconductor by external pressure



\bibitem{Bell} Ch.~J.~Raub, V.~B.~Compton, T.~H.~Geballe, B.~T.~Matthias, J.~P.~Maita, and G.~W.~Hull, Jr., J. Phys. Chem. Solids {\bf 26}, 2051 (1965).
%The occurence of superconductivity in sulfides, selenides, tellurides of Pt-group metals.



\bibitem{Canadian}
A.~Kjekshus, and W.~B.~Pearson, Can. J. Phys. {\bf 42}, 438 (1965).
%Constitution and Magnetic and Electrical properties of Palladium Tellurides (PdTe-PdTe2).

\bibitem{vandegrift} C. T. Van Degrift, Rev. Sci. Instrum. {\bf 46}, 599 (1975).



\bibitem{type1SC} H. Leng, C. Paulsen, Y. K. Huang, and A. de Visser, Phys. Rev. B {\bf 96}, 220506(R) (2017).
%Type-I superconductivity in the Dirac semimetal PdTe2. Unusual surface state SC observed.

\bibitem{STM} S. Das, Amit, A. Sirohi, L. Yadav, S. Gayen, Y. Singh, and G. Sheet, Phys. Rev. B {\bf 97}, 014523 (2018).
%Conventional superconductivity in the type-II Dirac semimetal PdTe2

\bibitem{HC}
Amit and Y. Singh, arXiv.1801.03288 (2018).
%Heat capacity evidence for conventional superconductivity in the Type-II Dirac semi-metal PdTe2





\bibitem{Canfield2016}
P.~C.~Canfield, T.~Kong, U.~S.~Kaluarachchi, and N.~H.~Jo, Phil. Mag. {\bf 96}, 84 (2016).
%Use of frit-disc crucibles for routine and exploratory solution growth of single crystalline samples

\bibitem{Canfield1992} P.~C.~Canfield and Z.~Fisk, Phil. Mag. B  {\bf 65}, 1117 (1992).
%GROWTH OF SINGLE-CRYSTALS FROM METALLIC FLUXES

\bibitem{FeSedetwinning} M. A. Tanatar, A. E. B\"ohmer, E. I. Timmons, M. Schütt, G. Drachuck, V. Taufour, K. Kothapalli, A. Kreyssig, S. L. Bud’ko, P. C. Canfield, R. M. Fernandes, and R. Prozorov, Phys. Rev. Lett. {\bf 117}, 127001 (2016).

\bibitem{Montgomery}
H. C. Montgomery, J. Appl. Phys. {\bf 42}, 2971 (1971).
\bibitem{Montgomery2}
B. F. Logan, S. O. Rice, and R. F. Wick, J. Appl. Phys. {\bf 42}, 2975 (1971).


\bibitem{Prozorov2000} R. Prozorov, R. W. Giannetta, A. Carrington, and F. M. Araujo-Moreira, Phys. Rev. B {\bf 62}, 115 (2000).
%Meissner-London state in superconductors of rectangular cross section in a perpendicular magnetic field

\bibitem{Prozorov2006} R. Prozorov and R. W. Giannetta, Supercond. Sci. Technol. {\bf 19}, R41 (2006).

\bibitem{MTC} M.~A.Tanatar, V.~A.~Bondarenko, E.~I.~Timmons, and R.~Prozorov, Rev. Sci. Instr. {\bf 89}, 013903 (2018).

\bibitem{Debay} K. Kudo, H. Ishii, and M. Nohara,
Phys. Rev. B {\bf 93}, 140505(R) (2016).
%Composition--induced structural instability and strong-coupling superconductivity in Au1−xPdxTe2

\bibitem{FS} J.~P.~Jan and H.~L.~Skriver, J. Phys. F. Metal Phys. {\bf 7}, 1719 (1977).
%RELATIVISTIC BANDSTRUCTURE AND FERMI-SURFACE OF PDTE2 BY LMTO METHOD

\bibitem{ProzorovKogan2011} R. Prozorov and V. G. Kogan,
Rep. Progr. Phys., {\bf 74}, 124505 (2011).
%London penetration depth in iron-based superconductors,


\bibitem{BCS} J. R. Schrieffer, \emph{Theory of Superconductivity}, Westview Press (1971).

\bibitem{WF}
J. Paglione, M. A. Tanatar, D. G. Hawthorn, F. Ronning,
R.W. Hill, M. Sutherland, L. Taillefer, and C. Petrovic,
Phys. Rev. Lett. {\bf 97}, 106606 (2006).
%Nonvanishing Energy Scales at the Quantum Critical Point of CeCoIn5

\bibitem{thermalcond1stvs2nd}
J. Paglione, M. A. Tanatar, J.-Ph. Reid, H. Shakeripour, C. Petrovic, and L. Taillefer,
Phys. Rev. Lett. {\bf 117}, 016601 (2016)
%Quantum Critical Quasiparticle Scattering within the Superconducting State of CeCoIn5




\bibitem{Comment} The $e_{u(1,i)}$ state is a special case as it is nonunitary and, hence, leads to two gaps on an initially spin-degenerate Fermi surface \cite{SigristUeda}. While one of the gaps is non-zero, the other one necessarily has nodal points on the $k_z$ axis.

\bibitem{SigristUeda} M. Sigrist and K. Ueda, Rev. Mod. Phys. {\bf 63}, 239 (1991).

\bibitem{Invariants} X.-L. Qi, T. L. Hughes, and S.-C. Zhang, Phys. Rev. B {\bf 81}, 134508 (2010).

\bibitem{Volovik} G.  E.  Volovik, \textit{The  Universe  in  a  Helium  Droplet}, Oxford University Press, Oxford (2003).

\bibitem{Brydon} P. M. R. Brydon, S. Das Sarma, H.-Y. Hui, and J. D. Sau, Phys. Rev. B \textbf{90}, 184512 (2014).

\bibitem{Scheurer-2016} M. S. Scheurer, Phys. Rev. B \textbf{93}, 174509 (2016).

\end{thebibliography}
\end{document}